\documentclass[twoside,twocolumn,american,english]{revtex4}
\usepackage[T1]{fontenc}
\usepackage[latin9]{inputenc}
\usepackage{color}
\usepackage{amsmath}

\makeatletter
\@ifundefined{textcolor}{}
{%
 \definecolor{BLACK}{gray}{0}
 \definecolor{WHITE}{gray}{1}
 \definecolor{RED}{rgb}{1,0,0}
 \definecolor{GREEN}{rgb}{0,1,0}
 \definecolor{BLUE}{rgb}{0,0,1}
 \definecolor{CYAN}{cmyk}{1,0,0,0}
 \definecolor{MAGENTA}{cmyk}{0,1,0,0}
 \definecolor{YELLOW}{cmyk}{0,0,1,0}
}

\usepackage{babel}

\makeatother

\usepackage{babel}
\begin{document}

\title{Analytic second-order energy derivatives in natural orbital functional
theory}

\author{Ion Mitxelena$^{1}$ and Mario Piris$^{1,2}$ }

\address{$^{1}$Kimika Fakultatea, Euskal Herriko Unibertsitatea (UPV/EHU),
20080 Donostia, Euskadi, Spain, and Donostia International Physics
Center (DIPC), 20018 Donostia, Euskadi, Spain.\\
$^{2}$IKERBASQUE, Basque Foundation for Science, 48013 Bilbao, Euskadi,
Spain.\bigskip{}
 }
\begin{abstract}
The analytic energy gradients in the atomic orbital representation
have recently been published (J. Chem. Phys. 146, 014102, 2017) within
the framework of the natural orbital functional theory (NOFT). We
provide here an alternative expression for them in terms of natural
orbitals, and use it to derive the analytic second-order energy derivatives
with respect to nuclear displacements in the NOFT. The computational
burden is shifted to the calculation of perturbed natural orbitals
and occupancies, since a set of linear coupled-perturbed equations
obtained from the variational Euler equations must be solved to attain
the analytic Hessian at the perturbed geometry. The linear response
of both natural orbitals and occupation numbers to nuclear geometry
displacements need only specify the reconstruction of the second-order
reduced density matrix in terms of occupation numbers.
\end{abstract}
\maketitle

\section{Introduction}

The matrix of second-order energy derivatives with respect to nuclear
displacements, or just the Hessian, is directly related to many properties
of great interest to chemists \cite{Papai1990,Frisch1990,Thomas1993,Wong1996,Pulay2013}.
Derivative methods are widely used to characterize the stationary
points on the potential energy surface, but are also essential for
the study of high-resolution molecular spectroscopy \cite{Yamaguchi2011},
or geometry dependent molecular properties such as electrostatic moments
\cite{Mitxelena}.\textcolor{black}{{} Analytic first-order derivatives
for reduced density matrix (RDM) methods are well-established, e.g.
for the parametric second-order RDM method \cite{referee1-4}, as
well as} analytical expressions of second-order energy derivatives
are well-known for standard electronic structure methods. Nevertheless,
the latter are still missing for methods that have been appeared in
the last few decades, such as those derived directly from RDMs \cite{Mazziotti2007,Sokolovdcft,Piris2014a,referee1-1,referee1-2,referee1-3}
without using the wavefunction.

In fact, the Hamiltonian corresponding to Coulombic systems only involves
one- and two-particle operators, hence the ground-state energy of
an electronic system can be computed using the first- and second-order
RDMs, denoted hereafter as $\varGamma$ and $D$, respectively. Within
the Born-Oppenheimer approximation, the electronic energy is then
written as
\begin{equation}
E_{el}=\sum\limits _{ik}\varGamma_{ki}\mathcal{H}_{ki}+\sum\limits _{ijkl}D_{kl,ij}\left\langle ij|kl\right\rangle ,\label{Eelec_0}
\end{equation}
where $\mathcal{H}_{ki}$ are the one-electron matrix elements of
the core-Hamiltonian, whereas $\left\langle ij|kl\right\rangle $
are the two-electron integrals of the Coulomb interaction. 

Accordingly, the role of the N-particle wavefunction can be assumed
by RDMs. Of particular interest are one-particle theories, where the
ground-state energy is represented in terms of $\varGamma$, because
the necessary and sufficient conditions that guarantee the ensemble
N-representability of $\varGamma$ are well established and are very
easy to implement \cite{Coleman1963}. In addition, the unknown functional
in a $\varGamma$-based theory only needs to reconstruct the electron-electron
potential energy \cite{Piris2007}, which is a notable advantage over
the density functional theory, where the kinetic energy functional
needs also to be reconstructed. $\varGamma$-functional theories seem
a promising way of overcoming the drawbacks of density functional
approximations currently in use.

Most functionals employ the exact energy expression (\ref{Eelec_0})
but using solely a reconstruction functional $D\left[\varGamma\right]$.
This implies that the exact ground-state energy will not, in general,
be entirely rebuilt. Approximating the energy functional has important
consequences \cite{Piris2017}. First, the theorems obtained for the
exact functional $E\left[\varGamma\right]$ are no longer valid. The
point is that an approximate functional still depends on $D$. An
undesired implication of the $D$-dependence is that the functional
N-representability problem arises, that is, we have to comply the
requirement that $D$ reconstructed in terms of $\varGamma$ must
satisfy the same N-representability conditions \cite{Mazziotti2007,referee1-1}
as those imposed on unreconstructed second-order RDMs to ensure a
physical value of the approximate ground-state energy. Otherwise,
the functional approximation will not be correct since there will
not be an N-electron system with an energy value (\ref{Eelec_0}).
In addition, due to this $D$-dependence, the resulting functional
depends only implicitly on $\varGamma$ and is not invariant with
respect to a unitary transformation of the orbitals. 

Nowadays, the approximate functionals are constructed in the basis
where $\varGamma$ is diagonal, which is the definition of a natural
orbital functional (NOF). Accordingly, it is more appropriate to speak
of a NOF rather than a functional of $\varGamma$ due to the existing
dependence on $D$. In this vein, in the NOF theory (NOFT) \cite{Piris2007},
the natural orbitals (NOs) are the orbitals that diagonalize $\varGamma$
corresponding to an approximate energy expression, such as those obtained
from an approximate wavefunction. The electronic energy can therefore
be considered as a functional of the NOs and occupation numbers (ONs).
In the following, we refer only to this basis, hence the ground-state
functional for N-electron systems is given by the formula 
\begin{equation}
E_{el}=\sum_{i}n_{i}\mathcal{H}_{ii}+\sum\limits _{ijkl}D\left[n_{i},n_{j},n_{k},n_{l}\right]\left\langle ij|kl\right\rangle .\label{Eelec_1}
\end{equation}
In Eq. (\ref{Eelec_1}), $D\left[n_{i},n_{j},n_{k},n_{l}\right]$
represents the reconstructed two-particle RDM in terms of the ONs.
It is worth to note that we neglect any explicit dependence of $D$
on the NOs themselves because the energy functional has already a
strong dependence on the NOs via the two-electron integrals.

In the last two decades, much effort has been put into making NOFT
able to compete with well-established electronic structure methods
\cite{Piris2014a,Pernal2016}. In this vein, the analytic energy gradients
in the atomic orbital representation for NOFT were obtained recently
\cite{Mitxelena2017}. In the present paper, an alternative expression
for them in terms of the NOs is given. On the other hand, the analytical
calculation of second-order derivatives is also desirable over numerical
treatment when high accuracy is required. Here, for the first time
in the context of NOFT, the second-order analytic energy derivatives
with respect to nuclear displacements are given.

\section{The Hessian}

The procedure for the minimization of the energy (\ref{Eelec_1})
requires optimizing with respect to the ONs and the NOs, separately.
The method of Lagrange multipliers is used to ensure the orthonormality
requirement for the NOs, and the ensemble N-representability restrictions
on $\varGamma$, which reduce to $0\leq n_{i}\leq1$ and $\sum_{i}n_{i}=N$
\cite{Coleman1963}. The bounds on $\left\{ n_{i}\right\} $ are enforced
by means of auxiliary variables, so merely one Lagrange multiplier
$\mu$ is needed to assure normalization of ONs. Hence, the auxiliary
functional $\Lambda\left[\mathrm{N},\left\{ n_{i}\right\} ,\left\{ \phi_{i}\right\} \right]$
is given by 
\begin{equation}
\begin{array}{c}
\Lambda=E_{el}-\mu\left({\displaystyle \sum_{i}}n_{i}-N\right)-{\displaystyle \sum_{ki}}\lambda_{ik}\left(\left\langle \phi_{k}|\phi_{i}\right\rangle -\delta_{ki}\right).\end{array}\label{Lagrangian}
\end{equation}

By making (\ref{Lagrangian}) stationary with respect to the NOs and
ONs, we obtain the Euler equations: 
\begin{equation}
\frac{\partial E_{el}}{\partial n_{m}}=\mathcal{H}_{mm}+\sum\limits _{ijkl}\frac{\partial D_{kl,ij}}{\partial n_{m}}\left\langle ij|kl\right\rangle =\mu,\label{equation_for_occupations}
\end{equation}
\begin{equation}
\frac{\partial E_{el}}{\partial\phi_{m}^{*}}=n_{m}\hat{\mathcal{H}}\phi_{m}+\sum\limits _{ijkl}D_{kl,ij}\frac{\partial\left\langle ij|kl\right\rangle }{\partial\phi_{m}^{*}}=\sum_{k}\lambda_{km}\phi_{k}.\label{orbital_EULER_equation}
\end{equation}

Eq. (\ref{equation_for_occupations}) is obtained holding the orbitals
fixed, whereas the set of the orbital Euler Eqs. (\ref{orbital_EULER_equation})
is satisfied for a fixed set of occupancies. For the sake of simplicity,
we concern only on the use of real orbitals throughout this work.
At present, the procedure of solving simultaneously Eqs. (\ref{equation_for_occupations})
and (\ref{orbital_EULER_equation}) is carried out by the iterative
diagonalization method described in Ref. \cite{Piris2009a}, which
is based on the hermiticity of the matrix of Lagrange multipliers
$\lambda$ at the extremum, i.e. $\left[\lambda-\lambda^{\dagger},\varGamma\right]=0$
(where super-index $\dagger$ is used to express the conjugate transpose).

As it is shown in Ref. \cite{Mitxelena2017}, the first-order derivative
of the electronic energy with respect to Cartesian coordinate $x$
of nucleus $A$, written in the atomic orbital representation, reads
as 
\begin{equation}
\begin{array}{c}
{\displaystyle \frac{dE_{el}}{dx_{A}}}={\displaystyle \sum_{\mu\upsilon}}\Gamma_{\mu\upsilon}\dfrac{\partial\mathcal{H}_{\mu\upsilon}}{\partial x_{A}}+{\displaystyle \sum_{\mu\upsilon\eta\delta}}D_{\eta\delta,\mu\upsilon}\dfrac{\partial\left\langle \mu\upsilon|\eta\delta\right\rangle }{\partial x_{A}}\\
\\
-{\displaystyle \sum_{\mu\upsilon}}\lambda_{\mu\upsilon}\dfrac{\partial\mathcal{S_{\mu\upsilon}}}{\partial x_{A}},\qquad\qquad\qquad\qquad
\end{array}\label{eq:gradient_initial}
\end{equation}

so the energy gradient depends only on the explicit derivatives of
one- and two-electron integrals and the overlap matrix. Therefore,
there is no contribution from ONs, and the resulting Eq. (\ref{eq:gradient_initial})
does not require obtaining the NOs and ONs at the perturbed geometry.
One could differentiate Eq. (\ref{eq:gradient_initial}) to achieve
an expression for the Hessian, nevertheless, perturbation of both
NOs and ONs must be considered. For that purpose it is more convenient
to work in the natural orbital (NO) representation $\left\{ \phi_{i}\right\} $,
so that Eq. (\ref{eq:gradient_initial}) transforms into 
\begin{equation}
\begin{array}{c}
{\displaystyle \frac{dE_{el}}{dx_{A}}}={\displaystyle \sum_{i}}n_{i}{\displaystyle \frac{\partial\mathcal{H}_{ii}}{\partial x_{A}}}+{\displaystyle \sum_{ijkl}}D_{kl,ij}{\displaystyle \frac{\partial\left\langle ij|kl\right\rangle }{\partial x_{A}}}\qquad\qquad\\
\\
{\textstyle \;-{\displaystyle \sum_{ij}}S_{ij}^{x_{A}}\lambda_{ij},\quad S_{ij}^{x_{A}}={\displaystyle {\textstyle {\displaystyle \sum_{\mu\upsilon}}}}C_{\mu i}C_{\upsilon j}{\displaystyle \frac{\partial\mathcal{S_{\mu\upsilon}}}{\partial x_{A}}}}.
\end{array}\label{eq:gradient_MO}
\end{equation}
The NOs associated to the perturbed geometry are usually expressed
as a linear combination of those NOs corresponding to the reference
state, so a perturbation of $x_{A}$ up to first order will carry
out the next change in the $\phi_{i}$ 
\begin{equation}
\phi_{i}+\delta x_{A}\left(\sum_{j}U_{ij}^{x_{A}}\phi_{j}+\sum_{\mu}C_{\mu i}\frac{\partial\zeta_{\mu}}{\partial x_{A}}\right)+\mathcal{O}\left(\delta x_{A}^{2}\right).\label{eq:perturbation}
\end{equation}

In Eq. (\ref{eq:perturbation}), $\left\{ \zeta_{\mu}\right\} $ are
the atomic orbitals, whereas changes in NO coefficients are accounted
by standard coupled-perturbed coefficients $\left\{ U_{ij}^{x_{A}}\right\} $.

The orthonormality relation of the perturbed NOs provides the relationship
\cite{Yamaguchi2011} 
\begin{equation}
\frac{\partial S_{ij}}{\partial x_{A}}=U_{ij}^{x_{A}}+U_{ji}^{x_{A}}+S_{ij}^{x_{A}}=0,\label{U+U+S=00003D00003D0}
\end{equation}
which can be used to derive the relation 
\begin{equation}
\begin{array}{c}
{\displaystyle \sum_{ij}S_{ij}^{x_{A}}\lambda_{ij}=-2\sum_{ij}U_{ij}^{x_{A}}\lambda_{ij}},\end{array}\label{S-->U}
\end{equation}

so the electronic energy gradients with respect to Cartesian coordinate
$x$ of nucleus $A$ in the NO representation reads as 
\begin{equation}
{\displaystyle \begin{array}{c}
{\displaystyle \frac{dE_{el}}{dx_{A}}=}{\displaystyle \sum_{i}}n_{i}{\displaystyle \frac{\partial\mathcal{H}_{ii}}{\partial x_{A}}}+{\displaystyle \sum_{ijkl}}D_{kl,ij}{\displaystyle \frac{\partial\left\langle ij|kl\right\rangle }{\partial x_{A}}}\\
\\
+\;2{\displaystyle \sum_{ij}}U_{ij}^{x_{A}}\lambda_{ij}.\qquad\qquad\quad
\end{array}}\label{eq:gradient}
\end{equation}
We may obtain second derivatives of the NOF energy by differentiating
Eq. (\ref{eq:gradient}) with respect to coordinate $y$ of nucleus
$B$, namely, 
\begin{equation}
\begin{array}{c}
{\displaystyle \frac{d^{2}E_{el}}{dx_{A}dy_{B}}=\sum_{i}n_{i}\frac{\partial^{2}\mathcal{H}_{ii}}{\partial x_{A}\partial y_{B}}+\sum_{ijkl}D_{kl,ij}\frac{\partial^{2}\left\langle ij|kl\right\rangle }{\partial x_{A}\partial y_{B}}}\\
\\
{\displaystyle \qquad\quad\;+\,2\sum_{ij}U_{ij}^{y_{B}}\lambda_{ij}^{x_{A}}+2\sum_{ij}\frac{d}{dy_{B}}{\textstyle \left(U_{ij}^{x_{A}}\lambda_{ij}\right)}}\\
\\
{\displaystyle {\displaystyle \quad+\sum_{m}n_{m}^{y_{B}}\frac{\partial}{\partial n_{m}}\left(\frac{dE_{el}}{dx_{A}}\right)}.\qquad\quad}
\end{array}\label{eq: HESSIAN}
\end{equation}

The first two terms in Eq. (\ref{eq: HESSIAN}) contain the explicit
derivatives of the core Hamiltonian and the two-electron integrals,
respectively. The next two terms arise from the derivatives of NO
coefficients with respect to the nuclear perturbation. Finally, $n_{m}^{y_{B}}$
represents the change in ON $m$ due to perturbation $y_{B}$, so
the last term in Eq. (\ref{eq: HESSIAN}) accounts for the contribution
from the perturbation of the ONs.

Taking into account Eq. (\ref{orbital_EULER_equation}), the matrix
of Lagrange multipliers can be written as 
\begin{equation}
\lambda_{ij}=n_{j}\mathcal{H}_{ij}+2\,\sum_{mkl}D_{kl,jm}\left\langle im|kl\right\rangle ,\label{LAMBDA-1}
\end{equation}
so explicit derivatives read as 
\begin{equation}
\lambda_{ij}^{x_{A}}=n_{j}\frac{\partial\mathcal{H}_{ij}}{\partial x_{A}}+2\,\sum_{mkl}D_{kl,jm}\frac{\partial\left\langle im|kl\right\rangle }{\partial x_{A}}.\label{lambda-explicit}
\end{equation}

Regarding the fourth summation of Eq. (\ref{eq: HESSIAN}), a more
comprehensive expression can be obtained, namely, 
\begin{equation}
\begin{array}{c}
{\displaystyle {\textstyle {\displaystyle \sum_{ij}\frac{d}{dy_{B}}}\left(U_{ij}^{x_{A}}\lambda_{ij}\right)}=\sum_{ij}\left\{ \frac{dU_{ij}^{x_{A}}}{dy_{B}}\lambda_{ij}+U_{ij}^{x_{A}}\frac{d\lambda_{ij}}{dy_{B}}\right\} }\end{array},\label{eq:3rd summation}
\end{equation}
where the first term in Eq. (\ref{eq:3rd summation}) is given by
\cite{Yamaguchi2011} 
\begin{equation}
\frac{dU_{ij}^{x_{A}}}{dy_{B}}=U_{ij}^{x_{A}y_{B}}-\sum_{k}U_{ik}^{y_{B}}U_{kj}^{x_{A}}.\label{dU/dY}
\end{equation}
By using Eq. (\ref{U+U+S=00003D00003D0}) together with the orthonormality
relation of the NOs we arrive at \cite{Yamaguchi2011} 
\begin{equation}
\begin{array}{c}
{\displaystyle \frac{\partial^{2}S_{ij}}{\partial x_{A}\partial y_{B}}}=U_{ij}^{x_{A}y_{B}}+U_{ji}^{x_{A}y_{B}}-{\displaystyle \sum_{m}}\left\{ S_{im}^{y_{B}}S_{jm}^{x_{A}}+S_{jm}^{y_{B}}S_{im}^{x_{A}}\right.\\
\left.-U_{im}^{y_{B}}U_{jm}^{x_{A}}-U_{jm}^{y_{B}}U_{im}^{x_{A}}\right\} +{\displaystyle {\textstyle {\displaystyle \sum_{\mu\upsilon}}}}C_{\mu i}C_{\upsilon j}{\displaystyle \frac{\partial^{2}\mathcal{S_{\mu\upsilon}}}{\partial x_{A}\partial y_{B}}}=0,
\end{array}\label{eq:algebra_U_ab_S_ab=00003D00003D0}
\end{equation}
then

\begin{equation}
\begin{array}{c}
2{\displaystyle \sum_{ij}}U_{ij}^{x_{A}y_{B}}\lambda_{ij}={\displaystyle \sum_{ij}}\lambda_{ij}\left({\displaystyle \sum_{m}}\left\{ S_{im}^{y_{B}}S_{jm}^{x_{A}}+S_{jm}^{y_{B}}S_{im}^{x_{A}}\right.\right.\\
\left.\left.-U_{im}^{y_{B}}U_{jm}^{x_{A}}-U_{jm}^{y_{B}}U_{im}^{x_{A}}\right\} -{\displaystyle {\textstyle {\displaystyle \sum_{\mu\upsilon}}}}C_{\mu i}C_{\upsilon j}{\displaystyle \frac{\partial^{2}\mathcal{S_{\mu\upsilon}}}{\partial x_{A}\partial y_{B}}}\right).
\end{array}\label{eq:algebra_U_ab}
\end{equation}
The derivative of Lagrange multipliers is obtained differentiating
Eq. (\ref{LAMBDA-1}) 
\begin{equation}
\frac{d\lambda_{ij}}{dy_{B}}=\lambda_{ij}^{y_{B}}+\sum_{k}U_{ki}^{y_{B}}\lambda_{kj}+\sum_{kl}U_{kl}^{y_{B}}Y_{ijkl},\label{lambda-derivative}
\end{equation}
where 
\[
\begin{array}{c}
Y_{ijkl}=n_{j}\delta_{jl}\mathcal{H}_{ik}+2\,{\displaystyle \sum_{mn}}D_{ln,jm}\left\langle im|kn\right\rangle \\
+\;4\,{\displaystyle \sum_{mn}}D_{mn,jl}\left\langle ik|mn\right\rangle .\qquad
\end{array}
\]
In Eq. (\ref{lambda-derivative}), the response from ONs has been
omitted since it is included later. Overall the fourth summation in
Eq. (\ref{eq: HESSIAN}) is given by 
\begin{equation}
\begin{array}{c}
{\displaystyle \sum_{ij}\frac{d}{dy_{B}}{\textstyle \left(U_{ij}^{x_{A}}\lambda_{ij}\right)}=\sum_{ij}\biggl\{ U_{ij}^{x_{A}y_{B}}\lambda_{ij}+U_{ij}^{x_{A}}\lambda_{ij}^{y_{B}}}\\
\begin{array}{c}
\qquad\qquad{\displaystyle {\displaystyle \qquad\qquad\qquad+{\displaystyle {\displaystyle {\displaystyle \sum_{kl}U_{ij}^{x_{A}}}}}U_{kl}^{y_{B}}Y_{ijkl}\biggr\}}}\end{array}.
\end{array}\label{eq:3rd summation-1}
\end{equation}
In the last summation of Eq. (\ref{eq: HESSIAN}), the derivatives
with respect to the occupancies read as 
\begin{equation}
\begin{array}{c}
{\displaystyle {\displaystyle \frac{\partial}{\partial n_{m}}\left(\frac{\partial E_{el}}{\partial x_{A}}\right)=}\frac{\partial\mathcal{H}_{mm}}{\partial x_{A}}+2\sum_{ij}U_{ij}^{x_{A}}\frac{\partial\lambda_{ij}}{\partial n_{m}}}\\
\\
{\displaystyle \qquad\qquad\quad+\sum_{ijkl}\frac{\partial D_{kl,ij}}{\partial n_{m}}\frac{\partial\left\langle ij|kl\right\rangle }{\partial x_{A}}},
\end{array}\label{eq:contribution_occ}
\end{equation}
where 
\begin{equation}
\begin{array}{c}
{\displaystyle \frac{\partial\lambda_{ij}}{\partial n_{m}}=\delta_{mj}\mathcal{H}_{ij}+2\,\sum_{rkl}\frac{\partial D_{kl,jr}}{\partial n_{m}}\left\langle ir|kl\right\rangle }\end{array}.\label{eq:lambda-respect-OCC}
\end{equation}
Note that $\partial D_{kl,jr}/\partial n_{m}$ is determined by the
given two-particle RDM reconstruction $D\left[n_{i},n_{j},n_{k},n_{l}\right]$
(see Eq. \ref{Eelec_1}). Substituting Eqs. (\ref{eq:3rd summation-1})
and (\ref{eq:contribution_occ}) into Eq. (\ref{eq: HESSIAN}), we
obtain the general expression for the Hessian in the NO representation,
namely, 
\begin{equation}
\begin{array}{c}
{\displaystyle \frac{d^{2}E_{el}}{dx_{A}dy_{B}}=\sum_{i}n_{i}\frac{\partial^{2}\mathcal{H}_{ii}}{\partial x_{A}\partial y_{B}}+\sum_{ijkl}D_{kl,ij}\frac{\partial^{2}\left\langle ij|kl\right\rangle }{\partial x_{A}\partial y_{B}}}\\
\\
\qquad\quad\;+\;2\;{\displaystyle \sum_{ij}}\left(U_{ij}^{y_{B}}\lambda_{ij}^{x_{A}}+U_{ij}^{x_{A}}\lambda_{ij}^{y_{B}}+U_{ij}^{x_{A}y_{B}}\lambda_{ij}\right)\\
{\displaystyle \qquad\quad+\;2\,\sum_{ijkl}U_{ij}^{x_{A}}U_{kl}^{y_{B}}Y_{ijkl}+{\textstyle {\displaystyle {\displaystyle \sum_{m}n_{m}^{y_{B}}}\left(\frac{\partial\mathcal{H}_{mm}}{\partial x_{A}}\right.}}}\\
\qquad\;\,\left.{\displaystyle +\;2\sum_{ij}U_{ij}^{x_{A}}\frac{\partial\lambda_{ij}}{\partial n_{m}}+\sum_{ijkl}\frac{\partial D_{kl,ij}}{\partial n_{m}}\frac{\partial\left\langle ij|kl\right\rangle }{\partial x_{A}}}\right).
\end{array}\label{eq:Hessian_FINAL}
\end{equation}
In contrast to first-order energy derivatives, the calculation of
the analytic Hessian requires the knowledge of NOs and ONs at the
perturbed geometry, expressed in Eq. (\ref{eq:Hessian_FINAL}) by
coefficients $U$ and $n_{m}^{y_{B}}$, respectively. Both magnitudes
are obtained from the solution of coupled perturbed equations which
are the result of deriving the variational conditions (\ref{equation_for_occupations}-\ref{orbital_EULER_equation}).
It is worth noting that in the case of Eq. (\ref{orbital_EULER_equation}),
it is more convenient to use its combination with its Hermitian conjugate
equation that gives us the variational condition on the Hermiticity
of Lagrange multipliers ($\lambda-\lambda^{\dagger}=0$).

\section{Coupled-perturbed equations}

Coupled perturbed equations for NOs and ONs were derived by Pernal
and Baerends \cite{Pernal2006} to obtain the linear response of $\varGamma$
in a problem with a one-electron static perturbation in the Hamiltonian.
In particular, these equations were employed in the calculation of
the static polarizabilities of atoms and molecules. The formalism
was later extended by Giesbertz \cite{giesbertz_thesis} to deal with
pinned ONs.

Here we present the coupled perturbed equations for NOs and ONs considering
from the beginning that NOs have an explicit dependence on the perturbation
(Eq. \ref{eq:perturbation}) through the position dependence of the
basis functions. Therefore, instead of considering an anti-Hermitian
$U$ matrix as done in Refs. \cite{Pernal2006,giesbertz_thesis},
standard coupled-perturbed coefficients are related with the overlap
matrix $S$ by Eq. (\ref{U+U+S=00003D00003D0}). In addition, the
existence of a generalized Fock matrix has not been assumed in the
present derivation. Our coupled-perturbed equations are obtained from
the Euler equations (\ref{equation_for_occupations}-\ref{orbital_EULER_equation}),
which are valid for any approximate NOF.

For real orbitals, at the extremum, the total derivatives of the variational
condition on the Hermiticity of Lagrange multipliers vanishes, 
\begin{equation}
\frac{d}{dx_{A}}\left(\lambda_{ij}-\lambda_{ji}\right)=0.\label{eq:L-L}
\end{equation}
Taking into account Eqs. (\ref{lambda-derivative}) and (\ref{eq:lambda-respect-OCC}),
Eq. (\ref{eq:L-L}) can be rewritten as 
\begin{equation}
\begin{array}{c}
\lambda_{ij}^{x_{A}}-\lambda_{ji}^{x_{A}}+{\displaystyle \sum_{k}}\left(U_{ki}^{x_{A}}\lambda_{kj}-U_{kj}^{x_{A}}\lambda_{ki}\right)+{\displaystyle \sum_{kl}\left(U_{kl}^{x_{A}}\right.}\\
\\
\left.Y_{ijkl}-U_{kl}^{x_{A}}Y_{jikl}\right)+{\displaystyle \sum_{k}}{\displaystyle \left(\frac{\partial\lambda_{ij}}{\partial n_{k}}-\frac{\partial\lambda_{ji}}{\partial n_{k}}\right)}n_{k}^{x_{A}}=0.
\end{array}\label{eq:  lambda_a - lambda_a}
\end{equation}

Eq. (\ref{U+U+S=00003D00003D0}) can be used to simplify first and
second summations in Eq. (\ref{eq:  lambda_a - lambda_a}), namely,
\begin{equation}
{\displaystyle \begin{array}{c}
{\displaystyle \sum_{k}U_{ki}^{x_{A}}\lambda_{kj}={\displaystyle \sum_{k>l}}\left[U_{kl}^{x_{A}}\left(\lambda_{kj}\delta_{li}-\lambda_{lj}\delta_{ki}\right)\right.}\\
\\
\qquad\qquad\qquad\quad-\left.S_{kl}^{x_{A}}\lambda_{lj}\delta_{ki}\right]{\displaystyle -\frac{1}{2}\,\sum_{k}S_{kk}^{x_{A}}\lambda_{kj}\delta_{ki}},
\end{array}}\label{eq:algebra}
\end{equation}
\begin{equation}
\begin{array}{c}
{\displaystyle \sum_{kl}}U_{kl}^{x_{A}}Y_{ijkl}={\displaystyle \sum_{k>l}}\left[U_{kl}^{x_{A}}\left(Y_{ijkl}-Y_{ijlk}\right)\right.\\
\\
\begin{array}{c}
\qquad\qquad\qquad\quad-\left.S_{kl}^{x_{A}}Y_{ijlk}\right]{\displaystyle -\frac{1}{2}\,\sum_{k}S_{kk}^{x_{A}}Y_{ijkk}.}\end{array}
\end{array}\label{eq:algebra_2}
\end{equation}

Accordingly, Eq. (\ref{eq:  lambda_a - lambda_a}) can be rewritten
as

\begin{equation}
\begin{array}{c}
\lambda_{ij}^{x_{A}}-\lambda_{ji}^{x_{A}}+{\displaystyle \sum_{k}\left(\frac{\partial\lambda_{ij}}{\partial n_{k}}-\frac{\partial\lambda_{ji}}{\partial n_{k}}\right)}n_{k}^{x_{A}}\qquad\\
\\
-{\displaystyle \frac{1}{2}\,}{\displaystyle \sum_{k}}S_{kk}^{x_{A}}\left(\delta_{ki}\lambda_{kj}-\delta_{kj}\lambda_{ki}+Y_{ijkk}-Y_{jikk}\right)\\
\\
+\;{\displaystyle \sum_{k>l}}U_{kl}^{x_{A}}\left(\delta_{li}\lambda_{kj}-\delta_{ki}\lambda_{lj}-\delta_{lj}\lambda_{ki}+\delta_{kj}\lambda_{li}\right.\\
\qquad\quad\left.+Y_{ijkl}-Y_{ijlk}-Y_{jikl}+Y_{jilk}\right)\\
\\
{\displaystyle -{\displaystyle \sum_{k>l}}S_{kl}^{x_{A}}\left(\delta_{ki}\lambda_{lj}-\delta_{kj}\lambda_{li}+Y_{ijlk}-Y_{jilk}\right)}=0
\end{array}\label{eq:lambda-lamda_definitive}
\end{equation}
Let us now consider the Eq. (\ref{equation_for_occupations}) involving
derivatives with respect to ONs. A perturbation up to first order
transforms it into 
\begin{equation}
\begin{array}{c}
{\displaystyle \frac{\partial\mathcal{H}_{mm}}{\partial x_{A}}+{\displaystyle \sum_{ijkl}\frac{\partial D_{kl,ij}}{\partial n_{m}}\frac{\partial\left\langle ij|kl\right\rangle }{\partial x_{A}}}+}{\displaystyle \sum_{rijkl}\frac{\partial^{2}D_{kl,ij}}{\partial n_{m}\partial n_{r}}}\left\langle ij|kl\right\rangle n_{r}^{x_{A}}\\
\\
+\;2{\displaystyle \sum_{ij}}\left[U_{ij}^{x_{A}}\left(\delta_{jm}\mathcal{H}_{ij}+2{\displaystyle \sum_{rkl}\frac{\partial D_{kl,jr}}{\partial n_{m}}}\left\langle ir|kl\right\rangle \right)\right]=\mu^{x_{A}}.
\end{array}\label{eq:coupled_occ_perturbed}
\end{equation}
Taking into account Eq. (\ref{U+U+S=00003D00003D0}), Eq. (\ref{eq:coupled_occ_perturbed})
can be rewritten in compact form as 
\begin{equation}
{\displaystyle \sum_{r}W_{mr}n_{r}^{x_{A}}+\sum_{i>j}U_{ij}^{x_{A}}\left(E_{ij}^{m}-E_{ji}^{m}\right)=F_{m}^{x_{A}},}\label{eq:coupled_occ_response}
\end{equation}

where 
\[
\begin{array}{c}
F_{m}^{x_{A}}=\mu^{x_{A}}-{\textstyle \left({\displaystyle \frac{\partial\mathcal{H}_{mm}}{\partial x_{A}}}+{\displaystyle {\textstyle {\displaystyle \sum_{ijkl}\frac{\partial D_{kl,ij}}{\partial n_{m}}\frac{\partial\left\langle ij|kl\right\rangle }{\partial x_{A}}}}}\right)}\\
\\
+\;{\displaystyle \sum_{i>j}S_{ij}^{x_{A}}E_{ji}^{m}+{\textstyle \frac{1}{2}}\,\sum_{i}S_{ii}^{x_{A}}E_{ii}^{m}},\qquad\quad\\
\\
E_{ij}^{m}=2\delta_{jm}\mathcal{H}_{ij}+4{\displaystyle \sum_{rkl}\frac{\partial D_{kl,jr}}{\partial n_{m}}}\left\langle ir|kl\right\rangle ,\qquad\qquad\\
\\
W_{mr}={\displaystyle \sum_{ijkl}\frac{\partial^{2}D_{kl,ij}}{\partial n_{m}\partial n_{r}}}\left\langle ij|kl\right\rangle .\qquad\qquad\qquad\qquad
\end{array}
\]

Note that $E_{ij}^{m}$ relates to $\partial\lambda_{ij}/\partial n_{m}$
by a factor $1/2$ according to Eq. (\ref{eq:lambda-respect-OCC}),
so Eqs. (\ref{eq:lambda-lamda_definitive}) and (\ref{eq:coupled_occ_response})
can bring together to obtain the complete expression for the coupled-perturbed
NOF equations 
\begin{equation}
{\displaystyle \begin{array}{c}
{\displaystyle \forall_{i>j}\;\sum_{k>l}A_{ij,kl}U_{kl}^{x_{A}}+\left(E_{ij}^{k}-E_{ji}^{k}\right)n_{k}^{x_{A}}=B_{ij}^{x_{A}}}\\
\\
\quad{\displaystyle \forall_{i}\quad\sum_{k>l}\left(E_{kl}^{i}-E_{lk}^{i}\right)U_{kl}^{x_{A}}+W_{ik}n_{k}^{x_{A}}=F_{i}^{x_{A}}}
\end{array}}\label{eq:CP-NOF}
\end{equation}
where 
\[
\begin{array}{c}
{\displaystyle A_{ij,kl}=\delta_{li}\lambda_{kj}-\delta_{ki}\lambda_{lj}-\delta_{lj}\lambda_{ki}+\delta_{kj}\lambda_{li}\qquad}\\
\\
+\;Y_{ijkl}-Y_{ijlk}-Y_{jikl}+Y_{jilk},\quad
\end{array}
\]
\[
\begin{array}{c}
{\displaystyle B_{ij}^{x_{A}}=\sum_{k>l}S_{kl}^{x_{A}}\left(\delta_{ki}\lambda_{lj}-\delta_{kj}\lambda_{li}+Y_{ijkl}-Y_{jilk}\right)}\quad\\
\\
{\displaystyle \qquad\quad+\,\frac{1}{2}\,}{\displaystyle \sum_{k}}S_{kk}^{x_{A}}\left(\delta_{ki}\lambda_{kj}-\delta_{kj}\lambda_{ki}+Y_{ijkk}-Y_{jikk}\right)\\
\\
-\,\lambda_{ij}^{x_{A}}+\lambda_{ji}^{x_{A}}.\qquad\qquad\qquad\qquad\qquad\;
\end{array}
\]

It is worth noting that the coupled-perturbed equations given by Eq.
(\ref{eq:CP-NOF}) are totally general and can be easily implemented,
so that an expression for the reconstructed $D\left[n_{i},n_{j},n_{k},n_{l}\right]$
is only required. The here presented formulation of such equations
exploits Eq. (\ref{U+U+S=00003D00003D0}) to calculate only necessary
$U$ coefficients, namely, the lower (or upper) block of matrix $U$.

The matricial form of Eq. (\ref{eq:CP-NOF}) is 
\begin{equation}
\left(\begin{array}{cc}
A & E-E^{\dagger}\\
E-E^{\dagger} & W
\end{array}\right)\left(\begin{array}{c}
U^{x_{A}}\\
n^{x_{A}}
\end{array}\right)=\left(\begin{array}{c}
B^{x_{A}}\\
F^{x_{A}}
\end{array}\right),\label{eq:CP-NOF___MATRIX}
\end{equation}
where $E^{\dagger}$ represents conjugate transpose operation only
acting on the subindexes, and it makes clear the symmetric nature
of the square matrix. The latter has to be computed and inverted only
once, since it is independent of the perturbation $\delta x_{A}$,
and presents only dependence on non-perturbed NOs and ONs.\bigskip{}

\section{Closing remarks}

Simple analytic expressions have been derived for computation of the
second-order energy derivatives with respect to nuclear displacements
in the context of the natural orbital functional theory. An alternative
expression for analytic gradients in terms of the NOs is given as
well. In contrast to first-order energy derivatives, the calculation
of the analytic Hessian requires the knowledge at the perturbed geometry
of NOs and ONs, which are obtained from the solution of coupled-perturbed
equations.

The coupled-perturbed equations were obtained from the corresponding
variational Euler equations considering that also basis functions
have explicit dependence on the geometry perturbations. Consequently,
the linear response of both NOs and ONs to non-external perturbations
of the Hamiltonian, as in the case of nuclear geometry displacements,
can be easily obtained by solving a set of equations that only need
to specify the reconstruction of the second-order RDM in terms of
the ONs.

In geometry optimization problems, the algorithms that employ the
Hessian knowledge are superior with respect to methods that use only
the gradient. The Hessian can be used for the most efficient search
of an extremum, and to test whether an extremum is a minimum or maximum
too. The formulas here presented constitute the groundwork for practical
calculations related to second-order energy derivatives with respect
to nuclear displacements, such as computation of harmonic vibrational
frequencies and thermochemical analysis.

\subsection*{Acknowledgments}

Financial support comes from Eusko Jaurlaritza (Ref. IT588-13) and
Ministerio de Economia y Competitividad (Ref. CTQ2015-67608-P). \foreignlanguage{american}{One
of us (I.M.) is grateful to Vice-Rectory for research of the UPV/EHU
for the PhD. grant }(PIF//15/043\foreignlanguage{american}{).} The
SGI/IZO\textendash SGIker UPV/EHU is gratefully acknowledged for generous
allocation of computational resources.

\expandafter\ifx\csname natexlab\endcsname\relax\global\long\def\natexlab#1{#1}
\fi \expandafter\ifx\csname bibnamefont\endcsname\relax \global\long\def\bibnamefont#1{#1}
\fi \expandafter\ifx\csname bibfnamefont\endcsname\relax \global\long\def\bibfnamefont#1{#1}
\fi \expandafter\ifx\csname citenamefont\endcsname\relax \global\long\def\citenamefont#1{#1}
\fi \expandafter\ifx\csname url\endcsname\relax \global\long\def\url#1{\texttt{#1}}
\fi \expandafter\ifx\csname urlprefix\endcsname\relax\global\long\def\urlprefix{URL }
\fi \providecommand{\bibinfo}[2]{#2} \providecommand{\eprint}[2][]{\url{#2}}


\begin{thebibliography}{{citenamefont{{J. Russel Thomas, J. DeLeeuw Bradley, George Vacek, T.   Daniel Crawford}}(1993)}}
\bibitem[{citenamefont{Papai et~al.}(1990)citenamefont{Papai, St-Amant,   Ushio, and Salahub}}]{Papai1990}
\bibinfo{author}{\bibfnamefont{I.}~\bibnamefont{Papai}},
\bibinfo{author}{\bibfnamefont{A.}~\bibnamefont{St-Amant}},
\bibinfo{author}{\bibfnamefont{J.}~\bibnamefont{Ushio}},
\bibnamefont{and} \bibinfo{author}{\bibfnamefont{D.}~\bibnamefont{Salahub}},
\bibinfo{journal}{Int. J. Quantum Chem.} \textbf{\bibinfo{volume}{38}},
\bibinfo{pages}{29} (\bibinfo{year}{1990}).

\bibitem[{citenamefont{Frisch et~al.}(1990)citenamefont{Frisch, Head-Gordon,   and Pople}}]{Frisch1990}
\bibinfo{author}{\bibfnamefont{M.}~\bibnamefont{Frisch}},
\bibinfo{author}{\bibfnamefont{M.}~\bibnamefont{Head-Gordon}},
\bibnamefont{and} \bibinfo{author}{\bibfnamefont{J.}~\bibnamefont{Pople}},
\bibinfo{journal}{Chem. Phys. Lett.} \textbf{\bibinfo{volume}{141}},
\bibinfo{pages}{189} (\bibinfo{year}{1990}).

\bibitem[{citenamefont{{J. Russel Thomas, J. DeLeeuw Bradley, George Vacek, T.   Daniel Crawford}}(1993)}]{Thomas1993}
\bibinfo{author}{\bibnamefont{{J. Russel Thomas, J. DeLeeuw Bradley,
George Vacek, T. Daniel Crawford}}}, \bibinfo{journal}{J. Chem.
Phys.} \textbf{\bibinfo{volume}{99}}, \bibinfo{pages}{403}
(\bibinfo{year}{1993}).

\bibitem[{citenamefont{Wong}(1996)}]{Wong1996} \bibinfo{author}{\bibfnamefont{M.~W.}
\bibnamefont{Wong}}, \bibinfo{journal}{Chem. Phys. Lett.} \textbf{\bibinfo{volume}{256}},
\bibinfo{pages}{391} (\bibinfo{year}{1996}).

\bibitem[{citenamefont{Pulay}(2014)}]{Pulay2013} \bibinfo{author}{\bibfnamefont{P.}~\bibnamefont{Pulay}},
\bibinfo{journal}{WIREs Comput. Mol. Sci.} \textbf{\bibinfo{volume}{4}},
\bibinfo{pages}{169} (\bibinfo{year}{2014}).

\bibitem[{citenamefont{Yamaguchi and Schaefer}(2011)}]{Yamaguchi2011}
\bibinfo{author}{\bibfnamefont{Y.}~\bibnamefont{Yamaguchi}}
\bibnamefont{and} \bibinfo{author}{\bibfnamefont{H.~F.} \bibnamefont{Schaefer}},
\emph{\bibinfo{title}{{Analytic Derivative Methods in Molecular
Electronic Structure Theory : A New Dimension to Quantum Chemistry
and its Applications to Spectroscopy}}} (\bibinfo{publisher}{John
Wiley and Sons, LTD}, \bibinfo{year}{2011}).

\bibitem[{citenamefont{Mitxelena and Piris}(2016)}]{Mitxelena} \bibinfo{author}{\bibfnamefont{I.}~\bibnamefont{Mitxelena}}
\bibnamefont{and} \bibinfo{author}{\bibfnamefont{M.}~\bibnamefont{Piris}},
\bibinfo{journal}{J. Chem. Phys.} \textbf{\bibinfo{volume}{144}},
\bibinfo{pages}{204108} (\bibinfo{year}{2016}).

\bibitem[{citenamefont{Mazziotti}(2007)}]{Mazziotti2007} \bibinfo{author}{\bibfnamefont{D.~A.}
\bibnamefont{Mazziotti}}, in \emph{\bibinfo{booktitle}{Reduced-Density-Matrix
Mechanics: with applications to many-electron atoms and molecules}},
edited by \bibinfo{editor}{\bibfnamefont{D.~A.} \bibnamefont{Mazziotti}}
(\bibinfo{publisher}{John Wiley and Sons}, \bibinfo{address}{Hoboken,
New Jersey, USA}, \bibinfo{year}{2007}), chap.~\bibinfo{chapter}{3},
pp. \bibinfo{pages}{21\textendash 59}, \bibinfo{edition}{1st}
ed.

\bibitem[{citenamefont{Sokolov et~al.}(2012)citenamefont{Sokolov, Wilke,   Simmonett, and Schaefer}}]{Sokolovdcft}
\bibinfo{author}{\bibfnamefont{A.~Y.} \bibnamefont{Sokolov}},
\bibinfo{author}{\bibfnamefont{J.~J.} \bibnamefont{Wilke}},
\bibinfo{author}{\bibfnamefont{A.~C.} \bibnamefont{Simmonett}},
\bibnamefont{and} \bibinfo{author}{\bibfnamefont{H.~F.} \bibnamefont{Schaefer}},
\bibinfo{journal}{J. Chem. Phys.} \textbf{\bibinfo{volume}{137}}
(\bibinfo{year}{2012}).

\bibitem[{citenamefont{Piris and Ugalde}(2014), (and references   therein)}]{Piris2014a}
\bibinfo{author}{\bibfnamefont{M.}~\bibnamefont{Piris}}
\bibnamefont{and} \bibinfo{author}{\bibfnamefont{J.~M.} \bibnamefont{Ugalde}},
\bibinfo{journal}{Int. J. Quantum Chem.} \textbf{\bibinfo{volume}{114}},
\bibinfo{pages}{1169} (\bibinfo{year}{2014), (and references
therein}).

\bibitem[{citenamefont{Coleman}(1963)}]{Coleman1963} \bibinfo{author}{\bibfnamefont{A.~J.}
\bibnamefont{Coleman}}, \bibinfo{journal}{Rev. Mod. Phys.}
\textbf{\bibinfo{volume}{35}}, \bibinfo{pages}{668} (\bibinfo{year}{1963}).

\bibitem[{citenamefont{Piris}(2007)}]{Piris2007} \bibinfo{author}{\bibfnamefont{M.}~\bibnamefont{Piris}},
in \emph{\bibinfo{booktitle}{Reduced-Density-Matrix Mechanics: with
applications to many-electron atoms and molecules}}, edited by \bibinfo{editor}{\bibfnamefont{D.~A.}
\bibnamefont{Mazziotti}} (\bibinfo{publisher}{John Wiley and
Sons}, \bibinfo{address}{Hoboken, New Jersey, USA}, \bibinfo{year}{2007}),
chap.~\bibinfo{chapter}{14}, pp. \bibinfo{pages}{387\textendash 427}.

\bibitem[{citenamefont{Piris}(2017)}]{Piris2017} \bibinfo{author}{\bibfnamefont{M.}~\bibnamefont{Piris}},
in \emph{\bibinfo{booktitle}{Many-body approaches at different scales:
a tribute to N. H. March on the ocasion of his 90th birthday}}, edited
by \bibinfo{editor}{\bibfnamefont{G. G. N.} \bibnamefont{Angilella}
and \bibfnamefont{C.} \bibfnamefont{Amovilli}} (\bibinfo{publisher}{Springer},
\bibinfo{address}{New York, USA}, \bibinfo{year}{2017}), chap.~\bibinfo{chapter}{22},
pp. \bibinfo{pages}{231\textendash 247}.

\bibitem[{citenamefont{Pernal and Giesbertz}(2016), (and references   therein)}]{Pernal2016}
\bibinfo{author}{\bibfnamefont{K.}~\bibnamefont{Pernal}}
\bibnamefont{and} \bibinfo{author}{\bibfnamefont{K.~J.~H.}
\bibnamefont{Giesbertz}}, \bibinfo{journal}{Top Curr Chem.}
\textbf{\bibinfo{volume}{368}}, \bibinfo{pages}{125} (\bibinfo{year}{2016),
(and references therein}).

\bibitem[{citenamefont{Mitxelena and Piris}(2017)}]{Mitxelena2017}
\bibinfo{author}{\bibfnamefont{I.}~\bibnamefont{Mitxelena}}
\bibnamefont{and} \bibinfo{author}{\bibfnamefont{M.}~\bibnamefont{Piris}},
\bibinfo{journal}{J. Chem. Phys.} \textbf{\bibinfo{volume}{146}},
\bibinfo{pages}{014102} (\bibinfo{year}{2017}).

\bibitem[{citenamefont{Piris and Ugalde}(2009)}]{Piris2009a} \bibinfo{author}{\bibfnamefont{M.}~\bibnamefont{Piris}}
\bibnamefont{and} \bibinfo{author}{\bibfnamefont{J.~M.} \bibnamefont{Ugalde}},
\bibinfo{journal}{J. Comput. Chem.} \textbf{\bibinfo{volume}{30}},
\bibinfo{pages}{2078} (\bibinfo{year}{2009}).

\bibitem[{citenamefont{Pernal and Baerends}(2006)}]{Pernal2006} \bibinfo{author}{\bibfnamefont{K.}~\bibnamefont{Pernal}}
\bibnamefont{and} \bibinfo{author}{\bibfnamefont{E.~J.} \bibnamefont{Baerends}},
\bibinfo{journal}{J. Chem. Phys.} \textbf{\bibinfo{volume}{124}},
\bibinfo{pages}{014102} (\bibinfo{year}{2006}).

\bibitem[{citenamefont{Giesbertz}(2010)}]{giesbertz_thesis} \bibinfo{author}{\bibfnamefont{K.~J.~H.}
\bibnamefont{Giesbertz}}, Ph.D. thesis, \bibinfo{school}{Vrije
Universiteit}, \bibinfo{address}{Amsterdam, The Netherlands} (\bibinfo{year}{2010}).

\bibitem[{citenamefont{referee1-1}(2017)citenamefont{mazziotti}}]{referee1-1}
\bibinfo{author}{\bibfnamefont{David~A.} \bibnamefont{Mazziotti}},
\bibinfo{journal}{Phys. Rev. Lett.} \textbf{\bibinfo{volume}{117}},
\bibinfo{pages}{153001} (\bibinfo{year}{2016}).

\bibitem{referee1-2} \bibinfo{author}{\bibfnamefont{Anthony~W.}
\bibnamefont{Schlimgen}}, \bibinfo{author}{\bibfnamefont{Charles~W.}
\bibnamefont{Heaps}}, \bibinfo{author}{\bibfnamefont{David~A.}
\bibnamefont{Mazziotti}}, \bibinfo{journal}{J. Phys. Chem. Lett.},
\textbf{\bibinfo{volume}{7} }(\textbf{\bibinfo{number}{4}}),
\bibinfo{pages}{627-631} (\bibinfo{year}{2016}).

\bibitem{referee1-3} \bibinfo{author}{\bibfnamefont{Alexandra~R.}
\bibnamefont{McIsaac}}, \bibinfo{author}{\bibfnamefont{David~A.}
\bibnamefont{Mazziotti}}, \bibinfo{journal}{Phys. Chem. Chem.
Phys.}, \textbf{\bibinfo{volume}{19}}, \bibinfo{pages}{4656-4660}
(\bibinfo{year}{2017}).

\bibitem{referee1-4} \bibinfo{author}{\bibfnamefont{Andrew~J.~S.}
\bibnamefont{Valentine}}, \bibinfo{author}{\bibfnamefont{David~A.}
\bibnamefont{Mazziotti}}, \bibinfo{journal}{Chem. Phys. Lett.},
\textbf{\bibinfo{volume}{685}}, \bibinfo{pages}{300-304} (\bibinfo{year}{2017}).
\end{thebibliography}
\end{document}